\documentclass{INTERSPEECH2023}
\usepackage{subcaption}
\usepackage{graphicx}

\interspeechcameraready
\title{Classifying Dementia in the Presence of Depression: A Cross-Corpus Study}
\name{Franziska Braun$^1$, Sebastian P. Bayerl$^1$, Paula A. Pérez-Toro$^2$, Florian Hönig$^4$, Hartmut Lehfeld$^3$, Thomas Hillemacher$^3$, Elmar Nöth$^2$, Tobias Bocklet$^1$, Korbinian Riedhammer$^1$}
\address{
  $^1$Technische Hochschule Nürnberg, 
  $^2$Friedrich-Alexander-Universität Erlangen-Nürnberg, Germany\\
  $^3$Klinik für Psychiatrie und Psychotherapie, Universitätsklinik der Paracelsus Medizinischen Privatuniversität, Klinikum Nürnberg, Germany\\
  $^4$KST Institut GmbH, Bad Emstal, Germany}
\email{franziska.braun@th-nuernberg.de}

\begin{document}

\maketitle
 
\begin{abstract}
Automated dementia screening enables early detection and intervention, reducing costs to healthcare systems and increasing quality of life for those affected.
Depression has shared symptoms with dementia, adding complexity to diagnoses. 
The research focus so far has been on binary classification of dementia (DEM) and healthy controls (HC) using speech from picture description tests from a single dataset. 
In this work, we apply established baseline systems to discriminate cognitive impairment in speech from the semantic Verbal Fluency Test and the Boston Naming Test using text, audio and emotion embeddings in a 3-class classification problem (HC vs. MCI vs. DEM). 
We perform cross-corpus and mixed-corpus experiments on two independently recorded German datasets to investigate generalization to larger populations and different recording conditions. 
In a detailed error analysis, we look at depression as a secondary diagnosis to understand what our classifiers actually learn.
\end{abstract}

\section{Introduction}
In recent years, a great deal of work has been done on the automatic detection of dementia (DEM) based on standardized neuropsychological tests such as the Cookie Theft Picture Description Test (CTP).
This is not surprising, as Alzheimer's disease (AD), the diagnosis for 60-70\% of dementia cases, is the most common neurodegenerative disease worldwide -- trending upward with demographic change and the associated aging of society. 
Through early detection and intervention, automated dementia screening methods can potentially reduce costs to healthcare systems and increase quality of life for affected individuals and their caregivers. 
Furthermore they are cheap and non-invasive.

However, numerous secondary diagnoses in older age add complexity to the diagnostic process.
The most common co-occurring and interdependent diagnosis with dementia is depression.
On the one hand, depression increases the risk of developing dementia, and on the other hand, the risk of depressive disorders is also significantly increased in people with dementia. 
Depressive disorders additionally impair the cognition, daily living functions and social skills of dementia patients, making them appear even more ``demented''.
Furthermore, depression can also cause cognitive disorders in the first place; such so-called pseudo-dementia can be difficult to detect even for clinical experts.

We conduct our research on two neuropsychological tests for which there has been little to no work in the field of automated dementia screening, but which are widely used in clinical practice, namely the semantic Verbal Fluency Test (sVFT) and the Boston Naming Test (BNT).
Aiming to prove the generalizability of established systems for dementia detection, we investigate the cross-corpus performance of BERT, W2V2, and emotion embedding features using data from a single-center and a multi-center study.
We hope to reduce selection biases introduced by, for example, recording conditions and local dialects and accents.
Another aspect of this work is the 3-class classification of HC, MCI, and DEM\footnote{We refer to DEM if the underlying data contains cases not related to AD.}.
From previous research we know that the binary classification between healthy and demented speech works well across conditions and languages.
This is to be expected, since in most cases this distinction should be possible even for non-experts.
The greater challenge is the detection of mild cognitive impairment (MCI), which is the transitional state between HC and DEM.
This task enables early detection (HC vs. MCI) and monitoring of dementia (MCI vs. DEM).
In a detailed error analysis, we show the role of depression as a secondary diagnosis in automated dementia screening and what our systems actually learn.
Our contributions are:
\begin{itemize}
    \item We evaluate a set of baseline methods using two established cognitive tests and two independently recorded German datasets.
    \item We study cross-corpus and mixed-corpus performance.
    \item We perform a detailed error analysis to investigate whether our classifiers learn to differentiate diseases or rather detect symptoms shared by these diseases.
\end{itemize}

\section{Related Work}\label{sc:related}
Most previous work has focused on binary classification of HC and cognitive disorders, such as MCI or DEM, using a single data set; the ADReSS and ADReSSo challenges established benchmarks and could advance the state of the art \cite{adress20,adresso21}. 
These focused on two tasks, the prediction of diagnosis in terms of a binary classification of AD vs. HC, and a regression task to predict MMSE score, using speech recordings and transcripts (ADReSS) or speech only (ADReSSo). 
The data consisted of recordings of the CTP from the Dementia Bank's publicly available Pitt Corpus \cite{becker1994dementiabank}.
The ADReSSo challenge also included a disease progression task using sVFT data.
Best challenge submissions achieved high accuracies using ERNIE \cite{yuan_disfluencies_2020} and BERT \cite{syed_automated_2020,pan_using_2021} text embedding features.

Picture description tasks such as the CTP have been extensively studied and have already been shown to be suitable for automatic detection of AD.
However, in clinical practice, the CTP and many other standardized tests are often used as part of neuropsychological test batteries, which enable differentiated examination of dementia.
Thus, investigation of other standardized tests could help to provide a more complete picture.
König \textit{et al.} showed that it is possible to automate fine-grained analyses of sVFT data for the assessment of cognitive decline \cite{koenig18} and demonstrated that, by using vocal markers in binary classification of MCI vs. DEM, fluent speech tasks even outperformed other vocal cognitive tasks such as picture descriptions \cite{konig_use_2018}.
In our previous work, we presented a study examining a number of standardized dementia screening tests, including the sVFT and the BNT, and showed that transcript-based automated scoring \cite{braun22_interspeech} and automatic classification of cognitive impairment using W2V2 features \cite{braun_GoingCookieTheft_2022} on these tests is possible.

Other studies analyzed the automatic detection of MCI from HC using acoustic, lexical, linguistic, and textual features extracted from speech from neuropsychological screening \cite{calza_linguistic_2021} and unstructured conversations \cite{asgari_predicting_2017, nagumo_automatic_2020}. 
A Hungarian research team developed a set of temporal speech parameters and linguistic features demonstrating their applicability to MCI and AD detection using data from the Hungarian MCI-mAD database \cite{toth_speech_2018, gosztolya_identifying_2019}.
In their recent work, Vincze \textit{et al.} reported accuracy values of 68-70\% for 3-class classification (HC vs. MCI vs. AD) using linguistic features only \cite{vincze_linguistic_2022}.

Cross-corpus studies are a major gap in the field of dementia research, which may be a source of disparity between different studies.
De~la~Fuente~Garcia \textit{et al.} investigated cross-corpus feature learning between spontaneous speech from monologues and dialogues for a binary classification of HC and AD \cite{la_fuente_garcia_cross-corpus_2020}.
If the classifiers were selected independently of feature extraction, they could demonstrate the generalizability of their feature model.
There have been also cross-corpus studies in context of cross-language learning from English to German or Spanish and vice-versa \cite{ablimit_exploring_2022,perez-toro_alzheimers_2022}.
While results from \cite{ablimit_exploring_2022} were barely above chance level, \cite{perez-toro_alzheimers_2022} showed good cross-language performance for BERT and W2V2 models.

\section{Data}\label{sc:data}
\subsection{Nuremberg Single-Center Corpus (NSC)}
The single-center corpus we used is drawn from our ongoing study \cite{braun22_interspeech} and includes 160 German-speaking subjects (63 men, 97 women) aged 49 to 89 years ($\mu = 73.65 \pm $8.97).
All tests and corresponding recordings were performed in the Memory Clinic of the Department of Psychiatry and Psychotherapy of Nuremberg Hospital as part of a face-to-face dementia screening and monitoring procedure consisting of a history taking interview and two neuropsychological test batteries (SKT and CERAD-NB), including a total of 18 tests.
Labels were provided by clinical experts and based on the Global Deterioration Scale (GDS) \cite{reisberg_global_1982}, which describes the seven stages of cognitive function in people with primary degenerative dementia. 
The data consists of 29 HC subjects from stages for no and age-related memory impairment (1--2), 48 MCI subjects from MCI stage (3), and 83 DEM subjects from mild to moderate dementia stages (4--5).
Detailed psychological and medical expert diagnoses include cognitive impairment level, depression level, and genesis, taking into account pre-existing conditions, medical examinations, questionnaires, cognitive testing, and clinical history taking with patients and their caregivers.

\subsection{PARLO Multi-Center Corpus (PMC)}
The multi-center corpus consists of 205 German-speaking subjects (95 men, 110 women) aged between 55 to 87 years ($\mu = 69.97 \pm $9.65); it was provided by the PARLO Institute for Research and Teaching in Speech Therapy.
Seven different cognitive tests (story reading, story recall, sVFT, BNT, word repetition, picture description, picture recall) and corresponding recordings were collected at nine academic memory clinics all-over Germany using a custom iPad app and the same type of iPad in a clinical setting.
This study is a cross-sectional, open-label, controlled, parallel-group clinical study of patients with AD-related MCI and mild-to-moderate AD, and cognitively healthy controls of comparable age.
The data comprises 82 HC, 58 MCI and 65 AD subjects.

\subsection{Neuropsychological Tests}\label{sc:tests}
We selected two tests that overlap in the two datasets: the sVFT and the BNT. 
The task of the sVFT \cite{isaacs73} is to name as many items as possible from a semantic category under a time constraint.
In this case, the subject was asked to name as many different animals as possible within one minute; the number of animals named gives the test score. 
The sVFT measures the speed and ease of verbal production ability, semantic memory, linguistic ability, executive functions, and cognitive flexibility. 

In contrast, the BNT \cite{kaplan78} is a confrontation naming test where the subject is asked to name objects presented in the form of line drawings; the number of correctly named objects forms the score.
The objects' names are distinguished according to the frequency of their occurrence in the German language. 
This task measures linguistic ability, visual perception, and word finding.
Short forms of the BNT were used, each comprising 15 drawings; the two corpora used a different set and order of objects.

Note that the sVFT recordings are very similar in both datasets, containing one snippet per subject consisting of about one minute of mostly patient speech.
The BNT sessions, however, differ greatly due to the objects presented and the recording procedure.
For the NSC, patient speech was recorded during the execution of the entire BNT, with durations varying from around 0.5 to 5 minutes.
For the PMC, on the other hand, 15 individual short recordings for each presented word were concatenated in a pre-processing step.
Both corpora occasionally contain speech fragments of the interviewer.

\section{Method}\label{sc:method}

\subsection{Text Embeddings}
Bidirectional Encoder Representations from Transformers (BERT) is a pre-trained masked language model for natural language processing (NLP) \cite{devlin_bert_2019}. 
It employs a transformer-based architecture to process massive amounts of text data and discover the relationships between words in a sentence.
BERT models can be fine-tuned for a variety of NLP tasks, including question answering, sentiment analysis, and text classification. 
The model is pre-trained to predict masked-out words and next-sentence prediction. 
BERT has already been successfully employed for AD detection in previous studies \cite{balagopalan_BERTNotBERT_2020,perez-toro_InfluenceInterviewerAutomatic_2021}. 
For our experiments, we use automatic transcripts, generated using OpenAI's Whisper\footnote{\url{https://github.com/openai/whisper}} (language=``de'', beam\_size=5, no\_speech\_threshold=0.8), and a BERT ``base'' model pre-trained on around 12GB of German text data (Wiki, OpenLegalData, News). 
We obtain the embeddings by sum pooling the last hidden states of the model.
The model weights are open-source and can be obtained online\footnote{\url{https://huggingface.co/bert-base-german-cased}}.

\subsection{Audio Embeddings}

Wav2vec 2.0 (W2V2) is a neural network model based on the transformer architecture designed for learning speech representations from raw audio data; it is usually pre-trained with a large amount of unlabeled audio data.
The W2V2 model processes raw waveform by a convolutional feature extractor, followed by 12 contextualized transformer blocks that use self-attention to focus on the parts of the audio material that are relevant to the task at hand (W2V2-base).
It can be used as a feature extractor with or without adaptation.
The features extracted from the model have been successfully used for detection of cognitive impairment in previous work \cite{braun_GoingCookieTheft_2022,balagopalan_comparing_2021}.
The W2V2 features used in our experiments were extracted from a model fine-tuned on the Mozilla Foundation Common Voice 9.0 dataset; model weights are open-source and can be obtained online\footnote{\url{https://huggingface.co/oliverguhr/wav2vec2-base-german-cv9}}.
The model takes z-normalized waveform data as input and returns 768-dimensional speech representations after each transformer block, representing about 0.02 seconds of audio.
This yields $N = T/0.02 - 1$ vectors for the extraction context of $T$ (i.e., 449 vectors for an extraction context of 10 seconds).
To obtain the final embeddings, we compute the mean vector over all extracted feature vectors of a sample, analogous to mean pooling along the time dimension.

\subsection{Emotion Embeddings}
The ``Emotional Model of Pleasure, Arousal, and Dominance'' (PAD)~\cite{mehrabian_PleasurearousaldominanceGeneralFramework_1996} represents different emotions in a three-dimensional space, where they can be either pleasant-unpleasant (valence), calm-agitated (arousal), or dominant-submissive (dominance). 
The PAD embeddings used in this work aim to capture similar aspects related to emotions, mood, and affective states in DEM patients, since studies suggest that the reduced ability of emotional perception in AD caused by memory loss may induce the appearance of apathy and depression \cite{henry_EmotionExperienceExpression_2009,goodkind_EmotionRegulationDeficits_2010}.
These embeddings were previously used to detect AD and to model depression in patients suffering from Parkinson's disease \cite{perez-toro_InfluenceInterviewerAutomatic_2021,perez-toro_EmotionalStateModeling_2021}.
We extract the arousal embeddings (last hidden layer) from a model pre-trained on the IEMOCAP dataset \cite{perez-toro_InfluenceInterviewerAutomatic_2021}.

\section{Experiments}\label{sc:exp}
Our experiments aim to distinguish the speech of individuals with no cognitive impairment (HC) from the speech of individuals with MCI or dementia. 
Experiments are conducted with speech data from the two cognitive tests described in section~\ref{sc:tests} using the three embeddings from section~\ref{sc:method}.

Since they have already been shown to be suitable for our purposes in existing work \cite{braun_GoingCookieTheft_2022}, we use Support Vector Machine (SVM) classifiers with Radial Basis Function Kernel (rbf) and Linear Kernel, which can also learn from only a few samples.
For within-dataset classification, we used stratified five-fold cross-validation (5-fold CV) of non-overlapping speakers by splitting the data of each dataset (i.e., NSC or PMC) into five distinct training and test sets comprising ~80\% and ~20\% of the data, respectively.
For mixed data set classification, we again use stratified 5-fold CV of disjoint speakers, but this time both data sets are split (i.e., NSC and PMC) and their respective training and test sets are combined.
For cross-dataset classification, we use all data from one dataset as the training set and test them with all data from the other dataset (i.e., train on NSC and test on PMC and vice versa).
Using grid search in another stratified 5-fold CV on the respective training portion, we determine the optimal hyperparameters for the SVM and the respective input features for the SVM classifiers.
The kernel parameter $\gamma$ is selected from the set
$ \gamma \in \{10^{-k} \mid k = 1, \ldots, 5 \} \subset \mathbb{R}_{>0} $,
the regularization parameter $C$ is selected from
$C \in \{10^{k} \mid k = -1, \ldots, 3 \} \subset \mathbb{N} $.
For the W2V2 experiments, the extracted layer $L$ is selected from $L \in \{1, 2, \ldots, 12\}$.

We use unweighted average recall (UAR) to evaluate the 3-class classification results, which is the arithmetic average of recall of all classes and is therefore a good indicator of model performance since NSC is imbalanced.

\section{Results}\label{sec:results}
\tablename~\ref{tab:results} shows the 3-class classification (HC vs. MCI vs. DEM) results using BERT, W2V2 and PAD features extracted from speech of the BNT and sVFT data, respectively.
Within-corpus and mixed-corpus performance is reported as average UAR with standard deviation over the five folds, while cross-corpus performance is reported as UAR on the particular dataset used as test partition.

\begin{figure}[t]
  \centering
  \includegraphics[width=\linewidth]{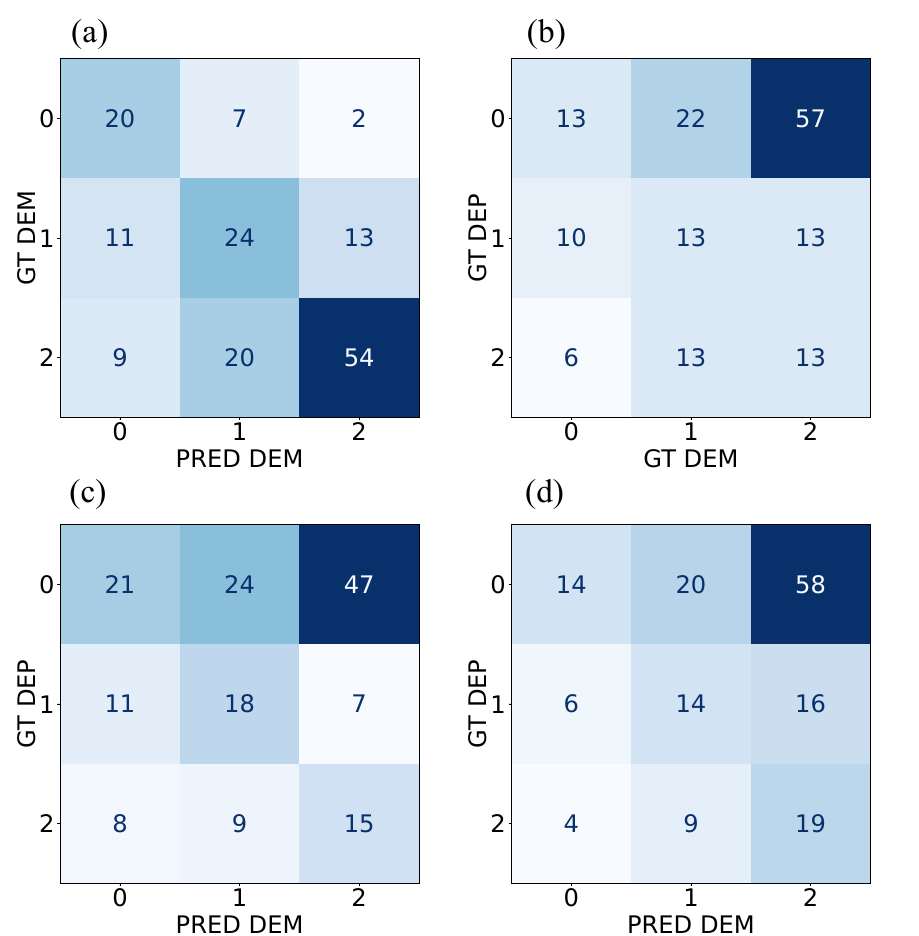}
  \caption{
  (a) Confusion matrix for ground truth (GT) and predicted (PRED) HC (0), MCI (1) and DEM (2);
  (b) GT Label co-occurrence of dementia and none (0), mild (1) and moderate-to-severe (3) depression;
  (c) Confusion matrix for NSC-trained system evaluated against depression;
  (d) Confusion matrix for PMC-trained system evaluated against depression.
  Reporting on W2V2 for sVFT; absolute numbers of NSC subjects.}
  \label{fig:conf_mat}
\end{figure}

For within- and mixed-corpus experiments, W2V2 consistently yields best results.
BERT is the overall runner-up, with the exception of BNT on NSC (within-corpus); this may be an artifact of the data since the mixed-corpus results are competitive.
\figurename~\ref{fig:conf_mat}(a) shows the confusion matrix for the W2V2 system on NSC for sVFT data.

For cross-corpus experiments, systems generally degrade as expected. 
However, BERT performs best on sVFT but worst on BNT. 
A likely explanation is that the system is able to focus on the actually spoken words; this is beneficial for sVFT but hurts for BNT, because of the test vocabulary mismatch between NSC and PMC.
The fact that the BERT sVFT systems actually perform better when tested on the \textit{other corpus} could be attributed to the larger amount of training data in the cross-corpus scenario.

PAD-based systems performance remains consistent across all setups, if the test data matches what has been observed in training: it drops severely for sVFT in cross-corpus experiments (NSC $\rightarrow$ PMC) but performs best in the mixed-corpus scenario; this can be explained with the diverse set of dialects and accents in PMC in contrast to the relatively homogeneous language in NSC.

\begin{table}[th]
\caption{Unweighted average recall for 3-class classification (in \%) using BERT, W2V2 and PAD embeddings for the sVFT and BNT on the NSC and PMC datasets.
For W2V2, we report best layer performance; for within-corpus and mixed-corpus combining both corpora (MIX), we use 5-fold cross-validation.
}
  \label{tab:results}
  \centering
\begin{tabular}{ cc|ccc }
\toprule
\textbf{train} & \textbf{test} & \textbf{BERT} & \textbf{W2V2} & \textbf{PAD}\\ 
\midrule
\multicolumn{5}{c}{\textbf{BNT}}                   \\
\midrule
NSC   & NSC  & 36.0$\pm$4.3 & 49.8$\pm$9.4  & 43.0$\pm$6.9\\
PMC   & PMC  & 49.2$\pm$4.8 & 56.3$\pm$12.3 & 39.9$\pm$7.1\\
\midrule
NSC   & PMC  & 34.4       & 41.9        & 37.4\\
PMC   & NSC  & 37.3       & 46.2        & 40.0\\
\midrule
MIX   & MIX  & 49.6$\pm$1.6 & 51.7$\pm$4.5 & 37.5$\pm$3.3\\
\midrule
\multicolumn{5}{c}{\textbf{sVFT}}                   \\
\midrule
NSC   & NSC  & 54.2$\pm$10.7 & 61.7$\pm$9.1 & 45.5$\pm$8.6\\ 
PMC   & PMC  & 46.8$\pm$8.9  & 55.4$\pm$6.7 & 44.2$\pm$10.6\\ 
\midrule
NSC   & PMC  & 49.3        & 43.8       & 35.4\\ 
PMC   & NSC  & 56.0        & 51.1       & 38.1\\ 
\midrule
MIX   & MIX  & 56.0$\pm$3.5  & 56.3$\pm$3.2 & 48.7$\pm$3.0\\
\bottomrule
\end{tabular}
\end{table}

\section{Error Analysis}
We performed an error analysis on NSC, since it provides detailed diagnoses by clinical experts.

The particularly bad performance of BERT for BNT when training or testing on NSC (rows 1, 3, 4 in Tab.~\ref{tab:results}) was in strong contrast to the mixed-corpus and overall sVFT performance.
An in-depth analysis of the transcripts revealed that the interviewer stated corrections for some of the more impaired patients, thus rendering those samples ill-suited for training or test.

Overall, if patients were misclassified from higher to lower cognitive impairment levels, about 83\% (BERT), 73\% (W2V2) and 29\% (PAD) of these subjects had above-average test performance, which was in stark contrast to their overall cognitive impairment level derived from the full assessment.
Conversely, below-average test performance resulted in a misclassification from lower to higher levels for 100\% (BERT), 44\% (W2V2) and 0\% (PAD) of such subjects.
This indicates that the BERT and W2V2 features tend to reflect test performance, while PAD features seem not to do so.  

Another challenge is the distinction between early dementia and MCI.
About 67\% (BERT), 60\% (W2V2) and 50\% (PAD) of the dementia subjects classified as MCI had mild or onset dementia.

We found that depression plays a major role in misclassification from lower to higher levels of cognitive impairment: about 68\% (BERT), 54\% (W2V2) and 47\% (PAD) of these misclassified patients suffered from depression.

This finding prompted us to investigate whether our classifiers learn diagnoses or rather \textit{symptoms} of diagnoses -- since dementia shares many symptoms with depression.
To this end, we evaluated our dementia classifiers trained on W2V2 on NSC or PMC by comparing their \textit{dementia predictions} on NSC against the \textit{depression labels}.
To do so, we group the depression levels in healthy, mild, and moderate-to-severe.
Note that these labels were inversely distributed, including 92 healthy, 36 mild, and 32 moderate-to-severe depressed subjects; \figurename~\ref{fig:conf_mat}(b) shows the label co-occurrence.


Following our initial experiments, we perform within- and cross-corpus evaluation, training on NSC and PMC, both testing on NSC; \figurename~\ref{fig:conf_mat}(c) and (d) show respective confusion matrices where we predict dementia but compare to depression.
If we were to classify symptoms shared by dementia and depression, the confusion matrices of (c) and (d) would feature a strong diagonal, implying that depression \textit{symptom} levels would be classified as respective dementia levels.
However, we find the confusion matrices are similar to the label co-occurrence in Fig.~\ref{fig:conf_mat}(b), suggesting that the classifiers capture actual dementia markers, thus making correct predictions for overlapping levels and incorrect predictions for non-overlapping levels.
It is striking that most subjects classified as demented have actually no symptoms of depression (cf. top right entry of (c) and (d)).
An instance intersection of (b) with (c) and (d), respectively, 
confirms that up to 81\% of the subjects from one matrix entry of (c) and (d) are contained in the same entry in (b).
For example, for no depression (0) and dementia (2), 46 subjects of (d) are contained in the 57 subjects of (b).

\section{Conclusion}
In this cross-corpus study, we investigated the generalization of three established baseline systems for automatic MCI and dementia detection using speech from two cognitive tests.
We showed that our systems are quite capable of generalizing and benefit from the increase in data.
We found that BERT features focus mostly on content (i.e., what is said), while W2V2 and PAD capture deeper speech features (i.e., how something is said).
Nevertheless, error analysis has shown that BERT and W2V2 features reflect \textit{test performance} as opposed to actual impairment level, while PAD features seem to ``look beyond'': even if overall the performance for PAD features was lower than with BERT and W2V2 features, they seem to carry other important discriminatory information. 

Detailed instance-based analyses showed that the dementia classifiers not only capture \textit{symptoms}, but can also provide diagnostic clues.
Preliminary experiments on training \textit{depression} classifiers on the NSC sessions resulted in close-to-chance performance, suggesting that the test design needs to match the diagnostic objective.

Future work on cross-test experiments and other related pathologies with common symptoms should be explored to confirm that test design and a combination of embeddings can form a valid diagnostic tool.
In the future, neural feature fusion techniques should be considered to combine the strengths of each feature to increase discriminatory power.



\newpage
\bibliographystyle{IEEEtran}
\footnotesize\bibliography{mybib,sb_extras}

\end{document}